\documentclass[preprints,article,accept,moreauthors,pdftex]{Definitions/mdpi} 
\firstpage{1} 
\pubvolume{1}
\issuenum{1}
\articlenumber{0}
\pubyear{2025}
\copyrightyear{2025}
\datereceived{ } 
\daterevised{ } 
\dateaccepted{ } 
\datepublished{ } 
\hreflink{https://doi.org/} 

\usepackage{amssymb}
\usepackage[labelformat=simple]{subcaption}
\renewcommand\thesubfigure{\alph{subfigure}}
\DeclareCaptionLabelFormat{subcaptionlabel}{\normalfont(\textbf{#2}\normalfont)}
\Title{Assessing Galaxy Rotation Kinematics: Insights from Convolutional Neural Networks on Velocity Variations}

\TitleCitation{Assessing Galaxy Rotation Kinematics using CNNs}

\Author{Amirmohammad Chegeni $^{1,2\ddagger}$\orcidA{}, Fatemeh Fazel Hesar $^{3,5\ddagger}$*\orcidB{}, Mojtaba Raouf $^{4,5,6}$\orcidC{}, Bernard Foing $^{4,5}$ and Fons J. Verbeek $^3$}

\AuthorNames{Amirmohammad Chegeni, Fatemeh Fazel Hesar, Mojtaba Raouf, Bernard Foing and Fons j. Verbeek}

\isAPAStyle{%
       \AuthorCitation{Lastname, F., Lastname, F., \& Lastname, F.}
         }{%
        \isChicagoStyle{%
        \AuthorCitation{Lastname, Firstname, Firstname Lastname, and Firstname Lastname.}
        }{
        \AuthorCitation{Lastname, F.; Lastname, F.; Lastname, F.}
        }
}

\address{%
$^{1}$ \quad Dipartimento di Fisica e Astronomia "G. Galilei", Università di
Padova, Via Marzolo 8, 35131 Padova, Italy; amirmohammad.chegeni@unipd.it\\
$^{2}$ \quad INFN-Padova, Via Marzolo 8, 35131 Padova, Italy; amirmohammad.chegeni@pd.infn.it\\
$^{3}$ \quad Leiden Institution of Advance Computer Science, 2300 RA Leiden, P.O. Box 9513, Netherlands; \\
$^{4}$ \quad Leiden Observatory, Leiden University, P.O. Box 9513, 2300 RA Leiden, Netherlands;\\

$^{5}$ \quad ILEWG LUNEX-EuroMoonMars Earth-Space, ESTEC European Space Agency, Keplerlaan 1, Noordwijk, 2201 AZ, Netherlands\\
$^{6}$ \quad School of Astronomy, Institute for Research in Fundamental Sciences (IPM), Tehran, 19395-5746, Iran}

\corres{Correspondence: f.fazel.hesar@liacs.leidenuniv.nl}

\firstnote{Current address: Affiliation.}  
\secondnote{These authors contributed equally to this work.}

\abstract{
Distinguishing galaxies as either fast or slow rotators plays a vital role in understanding the processes behind galaxy formation and evolution. 
Standard techniques, which are based on the $\lambda_R$-spin parameter obtained from stellar kinematics, frequently face difficulties to classify fast and slow rotators accurately. These challenges arise particularly in cases where galaxies have complex interaction histories or exhibit significant morphological diversity.
In this paper, we evaluate the performance of a Convolutional Neural Network (CNN) on classifying galaxy rotation kinematics based on stellar kinematic maps from the SAMI survey. Our results show that the optimal CNN architecture achieves an accuracy and precision of approximately 91\% and 95\% on the test dataset, respectively. Subsequently, we apply our trained model to classify previously unknown rotator galaxies for which traditional statistical tools have been unable to determine whether they exhibit fast or slow rotation, such as certain irregular galaxies or those in dense clusters. We also used Integrated Gradients (IG) to reveal the crucial kinematic features that influenced the CNN's classifications. This research highlights the power of CNNs to improve our comprehension of galaxy dynamics and emphasizes their potential to contribute to upcoming large-scale Integral Field Spectrograph (IFS) surveys.}


\begin{document}

\section{Introduction}

One of the key factors in investigating galaxy formation and evolution is defining the spin of the galaxy (rotation). As outlined by \cite{emsellem2007sauron}, the spin parameter ($\lambda_R$) provides a robust way to quantify the stellar angular momentum in galaxies. This parameter enables a clear distinction between fast and slow rotators, representing different evolutionary paths and formation mechanisms \cite{Raouf2021,Raouf2023,Raouf2024}. Fast rotators, characterized by high $\lambda_R$ values, are thought to have formed through gas-rich minor mergers or secular processes, including the inflow of external gas \citep{cappellari2007sauron}. In contrast, slow rotators, characterized by low $\lambda_R$ values, are believed to result from dissipationless processes, such as dry minor mergers \citep{hernquist1993structure, burkert2008sauron}. Furthermore, they note that the distribution of $\lambda_R$ values in different environments can provide insights into the role of galaxy interactions and environmental effects on galaxy evolution \citep{blanton2009physical, naab2014atlas3d,Raouf2014,Raouf2016,Raouf2018,Raouf2019}.

The methods for measuring galaxy spin have evolved significantly over the years, encompassing both observational and theoretical approaches. A significant advancement came with the introduction of the $V/\sigma$ parameter \citep{illingworth1977rotation}, which compared the ratio of rotation velocity to velocity dispersion and was introduced as an indicator of the observed rotation \citep{binney2005rotation}. In this method, integrated quantities such as $\langle V^2 \rangle$ and $\langle \sigma^2 \rangle$ are measured, representing the sky-averaged values of $V$ (velocity) and $\sigma$ (velocity dispersion)\footnote{The angle brackets indicate a sky-average weighted by surface brightness.}, as observed using integral-field spectrographs like SAURON \citep{emsellem2007sauron}. However, this method has limitations. For instance, galaxies such as NGC 3379, which exhibits global rotation, and NGC 5813, which shows spatially confined non-zero velocities, yield similar $V/\sigma$ values despite having distinct kinematic structures. On the other hand, the advent of integral-field spectroscopy, allowing for spatially resolved kinematics across entire galaxies. This led to the development of the $\lambda_R$ parameter by \cite{emsellem2007sauron}, which provided a more robust measure of specific angular momentum. 

The origin of angular momentum captured by the kinematic parameter $\lambda_R$, has been a topic of significant theoretical interest. For instance, \citep{vitvitska2002origin} propose an explanation for the origin of angular momentum in dark matter halos. On the theoretical front, numerical simulations have played a crucial role in understanding the origins and evolution of galaxy spin. 
In particular, \cite{prieto2015origin} conducted hydrodynamical simulations to show that the topology of the merging regions (specifically the number of intersecting filaments), accurately predicts the spin of both dark matter and gas.
They found that halos located at the centres of knots exhibit low spin, whereas those at the centres of filaments exhibit high spin.

Upcoming sky surveys, such as the Hector instrument \citep{bryant2020hector, bryant2024hector} and the Wide-field Spectroscopic Telescope (WST) \citep{Mainieri2024}, are expected to provide even larger databases with billions of galaxies. However, identifying slow/fast rotators within such an enormous number of galaxies cannot be achieved using traditional statistical methods and requires specialized tools. Convolutional neural networks (CNNs) \citep{a56}, a branch of machine learning models, have gained prominence in cosmology and astronomy. Their power lies in elucidating patterns within complex datasets, proving to be highly effective for diverse scientific tasks. In the field of exoplanet detection, CNN methods have been employed to enhance the process of identification and examination of potential exoplanets from large-scale datasets \citep{cnnast1, cnnast2}. In radio astronomy, deep learning approaches like DECORAS \citep{cnnast3} has been applied to differentiate and characterize radio-astronomical sources. They have also been helpful in Active galactic nuclei (AGNs) data \citep{GW1} such as detection and classification of  AGN host galaxies  \citep{chang2021identifying}. They have also been valuable in strong lensing searches \citep{st1,st2,st3}, improving efficiency by reducing the time required while enhancing accuracy. 
Furthermore, CNNs have proven to be valuable in cosmology,  showing potential to address computational limitations faced by conventional statistical methods in dark energy \citep{chegeni2023clusternets,escamilla2020deep,goh2024distinguishing}, dark matter \citep{cnnast7,lucie2019interpretable,herrero2022semi}, Cosmic Microwave Background (CMB) maps \citep{Sadr:2020rje,ni2023cmb,mishra2019cmb}, and Gravitational Wave (GW) \citep{GW,GW1,GW2}. These applications showcase the remarkable versatility and capability of CNNs n deepening our understanding of the universe.   \cite{REZAEI2025100921,baron2019machine,ntampaka2019role} provided an extensive overview of CNN applications in addressing a wide range of astronomical and cosmological challenges.

In this study, we leverage CNNs to classify galaxies as slow or fast rotators based on their stellar kinematics maps. We adopt a supervised learning approach, utilizing labeled data from the SAMI catalog as our training, testing, and validation datasets\footnote{We obtained results similar to those from the MaNGA dataset; however, we prefer to focus on the SAMI survey for this study, as it offers the advantage of higher signal-to-noise stellar kinematics in galaxies compared to MaNGA. While we did apply our method to a small sample of MaNGA data with known $\lambda$ and ellipticity values, the limited sample size and lower signal-to-noise ratio in MaNGA made it less valuable for inclusion at this stage.}. Through this process, we seek to find the best CNN architecture, achieving an accuracy of approximately 91\% on the test dataset. Subsequently, we apply our trained model to classify previously unknown rotators galaxies for which traditional statistical tools have been unable to determine whether they exhibit fast or slow rotation. We utilized interpretability techniques, including Integrated Gradients (IG), to uncover the key kinematic features that guided the CNN's classification decisions. Ultimately, this work highlights the effectiveness of CNNs in distinguishing between slow and fast rotators, demonstrating their potential to advance our understanding of galaxy dynamics and informing future observational strategies.
 
The structure of this paper is organized as follows. In Section \ref{sec2}, we describe the data source, preprocessing steps, and dataset preparation from the SAMI survey \citep{croom2012sydney,bryant2015sami}, which are used for the CNN-based classification of slow and fast galaxy rotators. In this section, we also provide a detailed explanation of the network architecture, data preprocessing, and evaluation methods. Our results and their physical interpretations are discussed in Section \ref{sec3}. In this Section, we also present a series of studies aimed at elucidating how our CNN model makes decisions when classifying images. Finally, we discuss and summarize the main results of this work in Section \ref{sec4}. 

\section{Dataset and Machine Learning Methods}\label{sec2}

The data source used in this study is the SAMI survey. The SAMI instrument \citep{croom2012sydney} was installed on the 3.9 meter Anglo-Australian Telescope (AAT) and connected to the AAOmega spectrograph \citep{sharp2006performance,
sharp2015sami,
bland2011hexabundles,
allen2015sami,
green2018sami,
scott2018sami}. This setup provides a median resolution of $\rm{FWHM}_{\rm{blue}} = 2.65 \AA$ in the range of $3700-5700 \textup{~\AA}$ and $\rm{FWHM}_{\rm{red}} = 1.61 \AA$ in the range of $6300-7400 \textup{~\AA}$ \citep{van2017sami}.
We concentrate on a sample that includes stellar kinematics, selecting only the Brightest Group Galaxies (BGGs) with stellar masses of $M_{\rm{star}} \gtrsim 10^{10.5} M_{\odot}$, ensuring the inclusion of massive galaxies with high signal-to-Noise (S/N $>$ 5).
In our study, we processed a dataset comprising 2,444 galaxies containing stellar kinematics data from the SAMI survey. We systematically iterated through the directory containing the galaxies FITS (Flexible Image Transport System) files, and extracting relevant information from each file. Specifically, we retrieved the primary data array from each FITS file and associated it with a unique identifier derived from the its filename. Then, we stored these data arrays and their corresponding identifiers in separate files.

For our data processing pipeline, we applied several crucial steps to ensure data quality and relevance for our analysis of galaxy stellar kinematics. First, we addressed the issue of duplicated labels in our dataset and removed any duplicate entries from our label array, ensuring each galaxy in our sample was represented uniquely. Thereafter, we handled missing values in our stellar kinematics data by replacing \textit{NaN} values with zeros, a common practice in astronomical data processing when dealing with regions of low signal-to-noise ratio. We categorised galaxies based on their rotational properties. Based on the spin-ellipticity relations from \citep{Emsellem2011}, we categorized each galaxy as either a fast rotator (designated as 1) or a slow rotator (designated as 0). Finally, to standardize our spatial analysis, we extracted a central region of each galaxy's kinematic map, creating a consistent spatial scale across our sample. By selecting a square region of $40 \times 40$ pixels centered on each galaxy, we ensured that our analysis focused on the most relevant and well-measured parts of each galaxy while maintaining a uniform spatial coverage across our sample. These data processing steps were crucial in preparing a clean, consistent, and well-characterized dataset for making our training, test and validation datasets. These datasets can used in our CNN binary classification problem. Then,
we applied a CNN to classify stellar kinematics maps that were previously labeled as fast or slow rotators.
\subsection{Convolutional Neural Network Architecture}
CNNs is a highly effective and widely used technique for object recognition and classification in the field of Machine Learning (ML) \citep{object}. These neural networks employ a hierarchical architecture consisting of multiple layers, including convolutional and pooling operations. Through this architecture, the CNN can automatically learn and extract meaningful features from input images, enabling it to process complex visual data with remarkable accuracy \citep{learning}. In our study, we applied a CNN on two-dimensional stellar kinematics maps to distinguish fast/slow rotators. 

We employed a systematic approach to design an optimal CNN architecture for classifying galaxies based on their stellar kinematic properties. We utilized \textit{Keras Tuner} \citep{kerastuner}, an automated hyperparameter optimization framework, to explore a wide range of model configurations efficiently. We varied the number of convolutional layers (between 1 and 4), the number of filters in each layer (from 32 to 256 in steps of 32), and the learning rate (choosing from 1e-2, 1e-3, or 1e-4). This approach enabled us to systematically evaluate different model complexities and training parameters.
The resulting optimal architecture, as determined by \textit{Keras Tuner} is showed in Table \ref{tab:cnn_architecture}.
\begin{table}[h]
\centering
\renewcommand{\arraystretch}{1.2}
\begin{tabular}{l c c}
\toprule
\textbf{Layer Type} & \textbf{Output Shape} & \textbf{Parameters} \\
\midrule
Conv2D & (None, 38, 38, 64) & 640 \\
MaxPooling2D & (None, 19, 19, 64) & 0 \\
BatchNormalization & (None, 19, 19, 64) & 256 \\
Conv2D & (None, 17, 17, 128) & 73,856 \\
MaxPooling2D & (None, 8, 8, 128) & 0 \\
BatchNormalization & (None, 8, 8, 128) & 512 \\
Conv2D & (None, 6, 6, 256) & 295,168 \\
MaxPooling2D & (None, 3, 3, 256) & 0 \\
BatchNormalization & (None, 3, 3, 256) & 1,024 \\
Flatten & (None, 2304) & 0 \\
Dense & (None, 96) & 221,280 \\
Dropout & (None, 96) & 0 \\
Dense & (None, 32) & 3,104 \\
Dropout & (None, 32) & 0 \\
Dense & (None, 1) & 33 \\
\bottomrule
\end{tabular}
\caption{Summary of the Convolutional Neural Network (CNN) architecture, including layer types, output shapes, and parameter counts.}
\label{tab:cnn_architecture}
\end{table}

In the CNN architecture, our goal is to minimise the binary cross-entropy loss. To achieve this, we employ the \textit{Adam} optimiser \citep{kingma2017adam} with a learning rate of 0.0002 and a beta parameter ($\beta_1$) of 0.9. Additionally, we implemented two checkpoints in the model. The first one is the \textit{ModelCheckpoint}, which saves the best model during training based on the validation accuracy. The second one is the \textit{ReduceLROnPlateau}, which reduces the learning rate when there is no progress in training. Moreover, we incorporate the \textit{EarlyStopping} callback, which stops the training process if the validation accuracy does not improve. To monitor the training progress, we recorded the loss function and accuracy computed on the training data after each epoch. We also calculate the validation loss and accuracy on a separate validation dataset. These metrics provide insights into the performance of the model during training and helped track its progress.

\subsection{Data Preprocessing}
We enhanced our dataset using a custom data augmentation function. This function takes the original data and labels as input and applies various transformations to generate additional training samples, thereby increasing the diversity of the training set. Specifically, we utilized the \textit{ImageDataGenerator} from \textit{Keras}, configured with parameters such as a rotation range of 360 degrees, width and height shifts of up to 4 pixels, zoom range from 1.0 to 1.05, and horizontal flipping, with the fill mode set to \textit{nearest} to maintain the integrity of the images. To balance the number of fast and slow rotator samples in the dataset, the function performed seven times more augmentations on the slow rotator samples. These transformed images were added to the dataset along with their corresponding labels, ensuring an equal representation of fast and slow rotators.

After preprocessing, the dataset consists of 1,112 samples, with 980 classified as fast rotators and 132 as slow rotators. To address the class imbalance, we applied data augmentation, specifically aimed at balancing the number of slow and fast rotators. As a result, the final dataset comprises 19,332 samples, with 9,603 fast rotators and 9,729 slow rotators, ensuring a more balanced distribution for training the model.
  Then we divided it into three subsets: 70\% for training, 15\% for testing, and the remaining for validation. This split ensured that the model was trained on a substantial portion of the data, while also being evaluated and validated on separate, unseen subsets to gauge its performance and generalization capabilities. This augmentation and splitting strategy was crucial for improving the robustness and accuracy of our CNN in classifying stellar kinematics maps as fast or slow rotators.

\subsection{Loss Function}
Loss functions are used to evaluate our network’s accuracy by measuring the inequality between predicted and true class labels. For our binary classification problem of distinguishing fast from slow rotators, we employed the Binary Cross Entropy (BCE) \citep{mao2023cross} loss function. Optimization was carried out using the \textit{Adam} optimizer with a learning rate of 0.001 and the BCE loss function is formulated as:

\[
{\rm BCE} = -\frac{1}{N} \sum_{i=1}^{N} \left[ y_i \log p_i + (1 - y_i) \log (1 - p_i) \right],
\]
In this context, \( y_i \) corresponds to the actual class label of the \( i \)th sample within our dataset of \( N \) training samples, and \( p_i \) is the probability predicted by the model for the \( i \)th sample being a fast or slow rotator.

\subsection{Evaluation Criteria}
To evaluate and compare the effectiveness of the CNN on the test dataset, it is crucial to establish appropriate evaluation criteria. In the context of stellar kinematics maps, which is treated as a classification problem distinguishing between fast/slow rotators samples, we can utilize the following confusion matrix for evaluation purposes:
\[
\begin{array}{cc|cc}
\multicolumn{2}{c}{} & \multicolumn{2}{c}{\text{True Data}} \\
\cline{3-4}
\multicolumn{2}{c|}{} & \text{Fast} & \text{Slow} \\
\cline{2-4}
\text{Test Results} & \text{Fast} & TP & FP \\
\cline{2-4}
 & \text{Slow} & FN & TN \\
\cline{2-4}
\end{array}
\label{tab:sample_confusion}
\]
In this context, True Positive (TP) indicates correctly identified fast rotator, True Negative (TN) corresponds to accurately recognized slow rotator, False Positive (FP) denotes misclassification of slow rotator sources as fast one, and False Negative (FN) refers to fast rotator missed by the algorithm and classified as slow one.

These terms allow us to define various evaluation metrics. Accuracy measures the fraction of correctly identified samples (TP and TN) out of the total number of samples. Precision quantifies the ratio of true positives to the sum of true positives and false positives, showing the reliability of positive predictions \citep{liang2022confusion}. These metrics are defined as:
\begin{equation}
\begin{aligned}
\text{Accuracy} &= \frac{TP + TN}{TP + FN + TN + FP} \\
\text{Precision} &= \frac{TP}{TP + FP}
\end{aligned}
\end{equation}
The Receiver Operating Characteristic (ROC) curve \citep{hanley1989receiver} is another useful metric, which depicts the trade-off between TP and the FP for the CNN model. Each point on the ROC curve corresponds to a different threshold for categorizing samples as slow or fast rotator based on their predicted probabilities. By analyzing the ROC curve, we can determine how well the model distinguishes between fast and slow rotator samples. A model with superior performance will have a curve that closely approaches the top-left corner of the plot,      signifying a higher TP rate and a lower FP rate for various thresholds.
\section{Results}\label{sec3}
\begin{figure}
		\centering
		\includegraphics[width=0.6\linewidth]{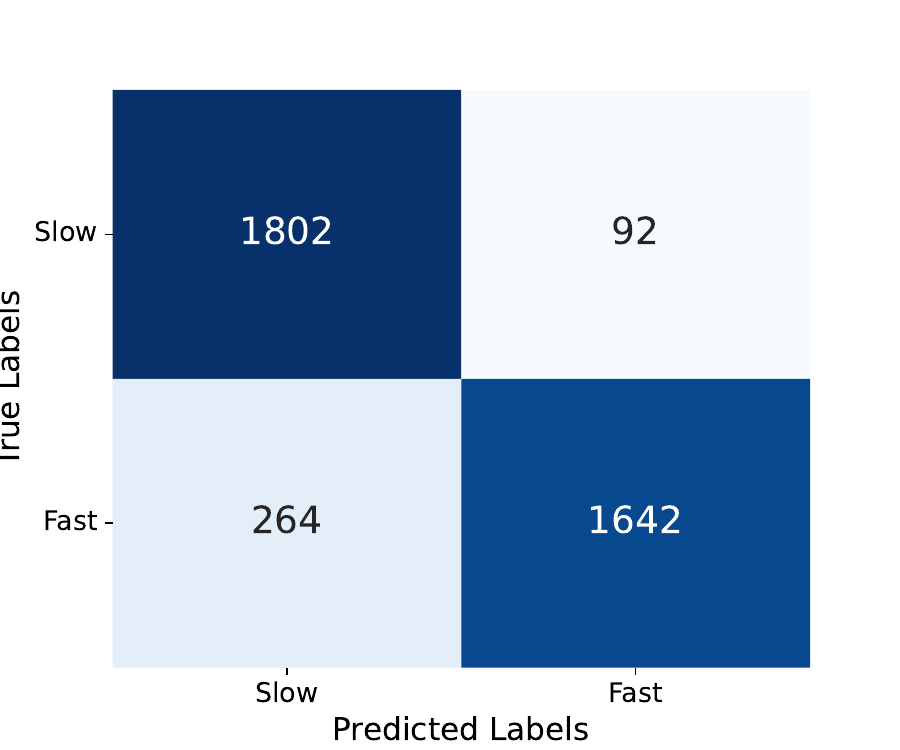}
		\caption{Confusion matrix illustrating model performance with an accuracy of 91\% and a precision of 95\%, highlighting its effectiveness in classifying slow and fast rotators.}
		\label{cm}
	\end{figure}
In this section, we present the results that show the performance of our CNN-based model in classifying fast and slow rotator using the stellar kinematics maps from the SAMI survey. We first investigate the model’s classification performance through the confusion matrix and the ROC curve, and analyse the distribution of the model’s predictions. Afterward, we apply the CNN model to unknown rotational samples and analyze the key features in the stellar maps that influence its decision-making process. To achieve this, we utilize an explainable CNN approach, including Integraded Gradients (IG).




  \begin{figure}
		\centering
		\includegraphics[width=\linewidth]{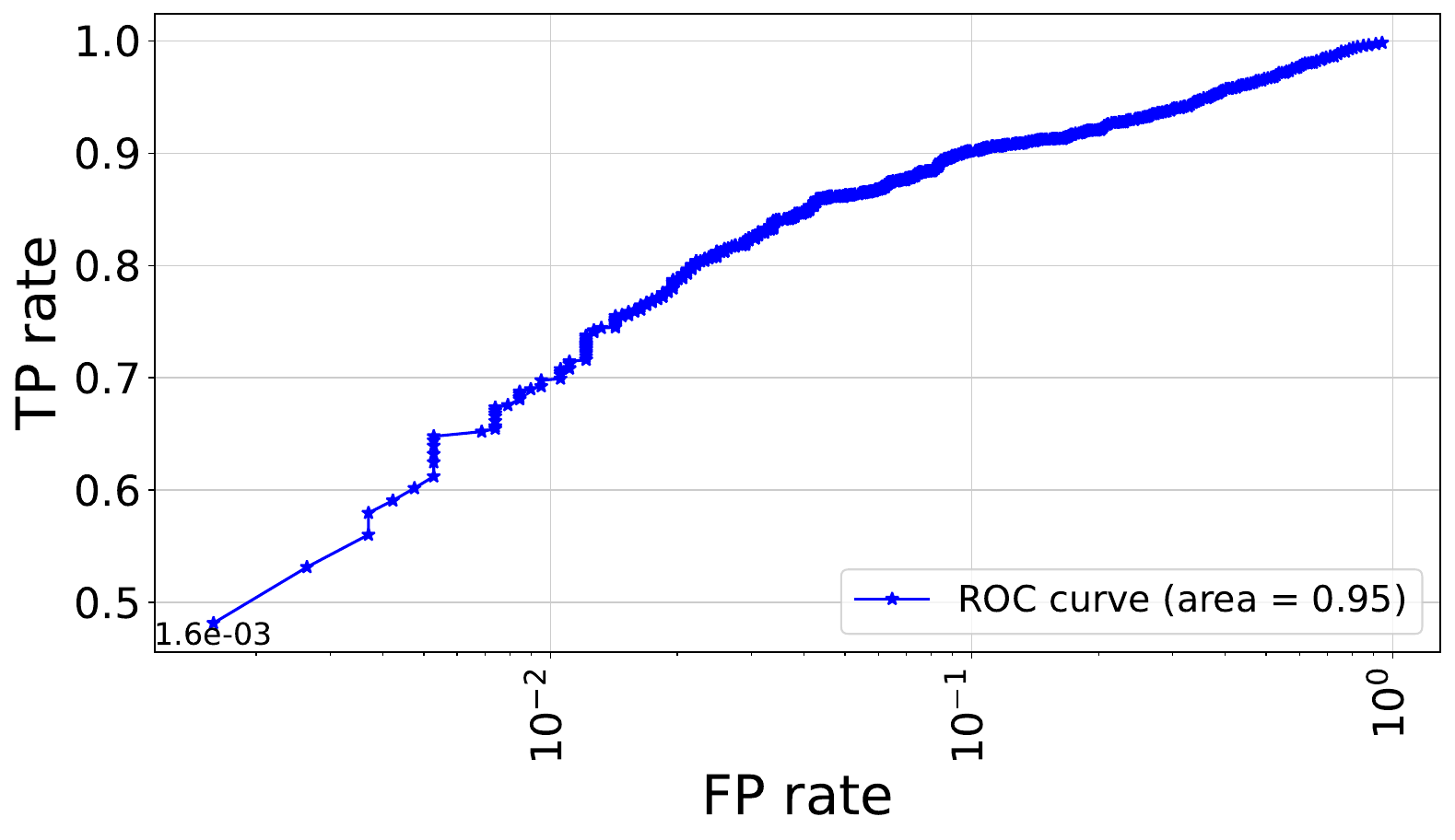}
		\caption{The ROC plot demonstrates how the CNN model perform in distinguishing between fast/slow rotators samples across various thresholds, balancing true positive (TP) and false positive (FP) rates. The plot uses a logarithmic scale for the x-axis (FP rate), allowing for a more detailed view of the model's performance at low FP rates}
		\label{roc}
	\end{figure}

   \begin{figure}
		\centering
		\includegraphics[width=\linewidth]{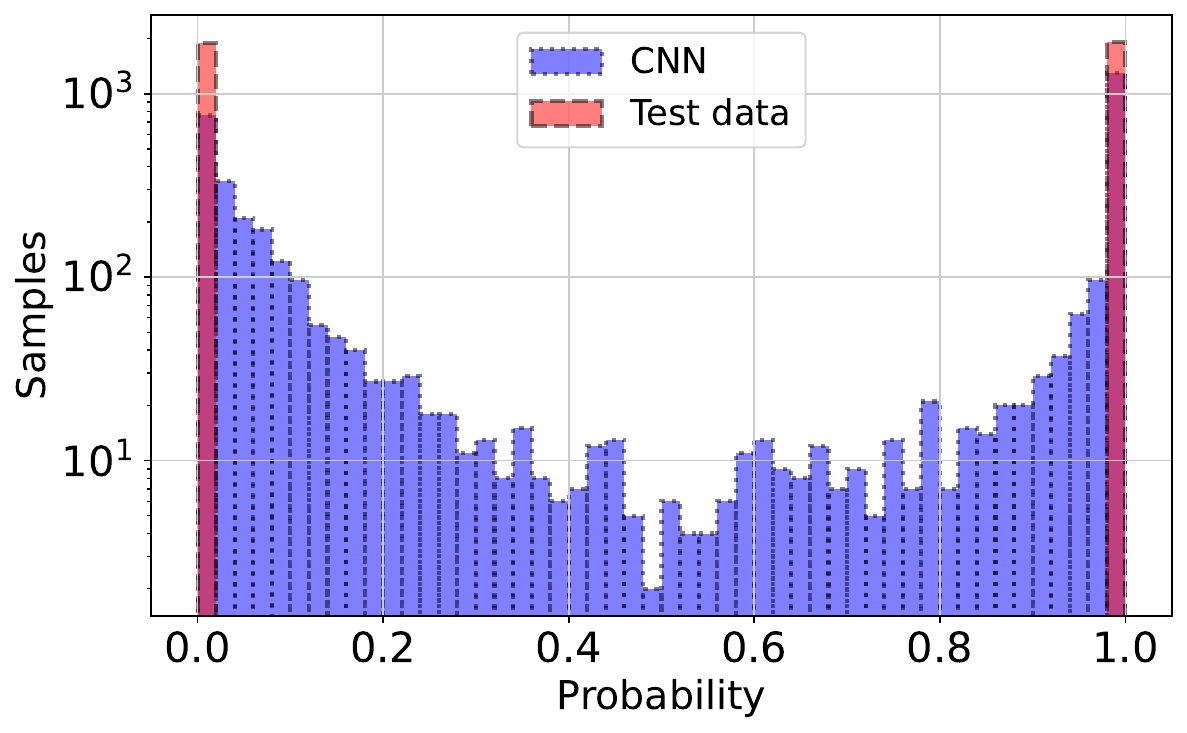}
		\caption{This distribution plot compares the predictions made by the CNN model against the real test data. The x-axis represents the prediction probability from 0 to 1, while the y-axis shows the number of samples on a logarithmic scale.}
		\label{prob}
	\end{figure}

\subsection{Training CNN on Galaxies with Known Fast/Slow Rotation}
Using SAMI observations, the stellar kinematics maps of galaxies can be estimated and the parameter $\lambda_R$ is used as a proxy for the
spin parameter.  Following \citep{emsellem2007sauron} it is
calculated using:
\begin{equation}
    \lambda_R = \frac{\sum_{i} (F_i \cdot R_i \cdot |V_i|)}{\sum_{i} (F_i \cdot R_i \cdot \sqrt{V_i^2 + \sigma_i^2})};
\end{equation}

\noindent where the subscript \(i\) refers to the \(i\)th spaxel within the ellipse, \(F_i\) is the flux of the \(i\)th spaxel, \(V_i\) is the stellar velocity in km s\(^{-1}\), and \(\sigma_i\) is the velocity dispersion in km s\(^{-1}\). We train the CNN  on a training data including stellar kinematics maps with the defined 7697 fast and 7835 slow rotators which defined by the parameter $\lambda_R$.

In Figure \ref{cm} we show the confusion matrix for the CNN model which is demonstrates a strong performance of the model, especially in terms of precision, suggesting that it effectively minimizes false positives for the slow rotator category. The overall accuracy further reinforces the model's reliability in making predictions between the two classes. Since the model output represents the probability of an
object being a fast rotator, ranging between 0 and 1, setting a threshold is
necessary to classify objects into two classes: fast and slow rotators. Figure \ref{roc}  illustrates the ROC plot which is followed a steep ascent, quickly reaching high TP rates even at low FP rates, which is indicative of a strong classifier. The Area Under the ROC Curve (AUC) is reported as 0.95, which is very close to the ideal value of 1.0. This high AUC score suggests that the model has excellent discriminative ability between the slow/fast rotator category. Figure \ref{prob}  reveals a distribution for both the CNN predictions (blue) and the real data (pink). There are two prominent peaks at the extreme ends (0 and 1) for both distributions, indicating that the model is making confident predictions for a large number of samples, classifying them as either strongly slow rotators (close to 0) or strongly fast rotators (close to 1).
The CNN's predictions closely align with the real data distribution, especially at the extremes. This suggests that the model is performing well in identifying clear-cut cases. However, there are some discrepancies in the middle range (around 0.2 to 0.8), where the CNN shows more varied predictions compared to the real data. This could indicate that the model is less certain about borderline cases and produces a wider range of probability estimates for these samples.

\begin{figure*}
		\centering
		\includegraphics[width=\linewidth]{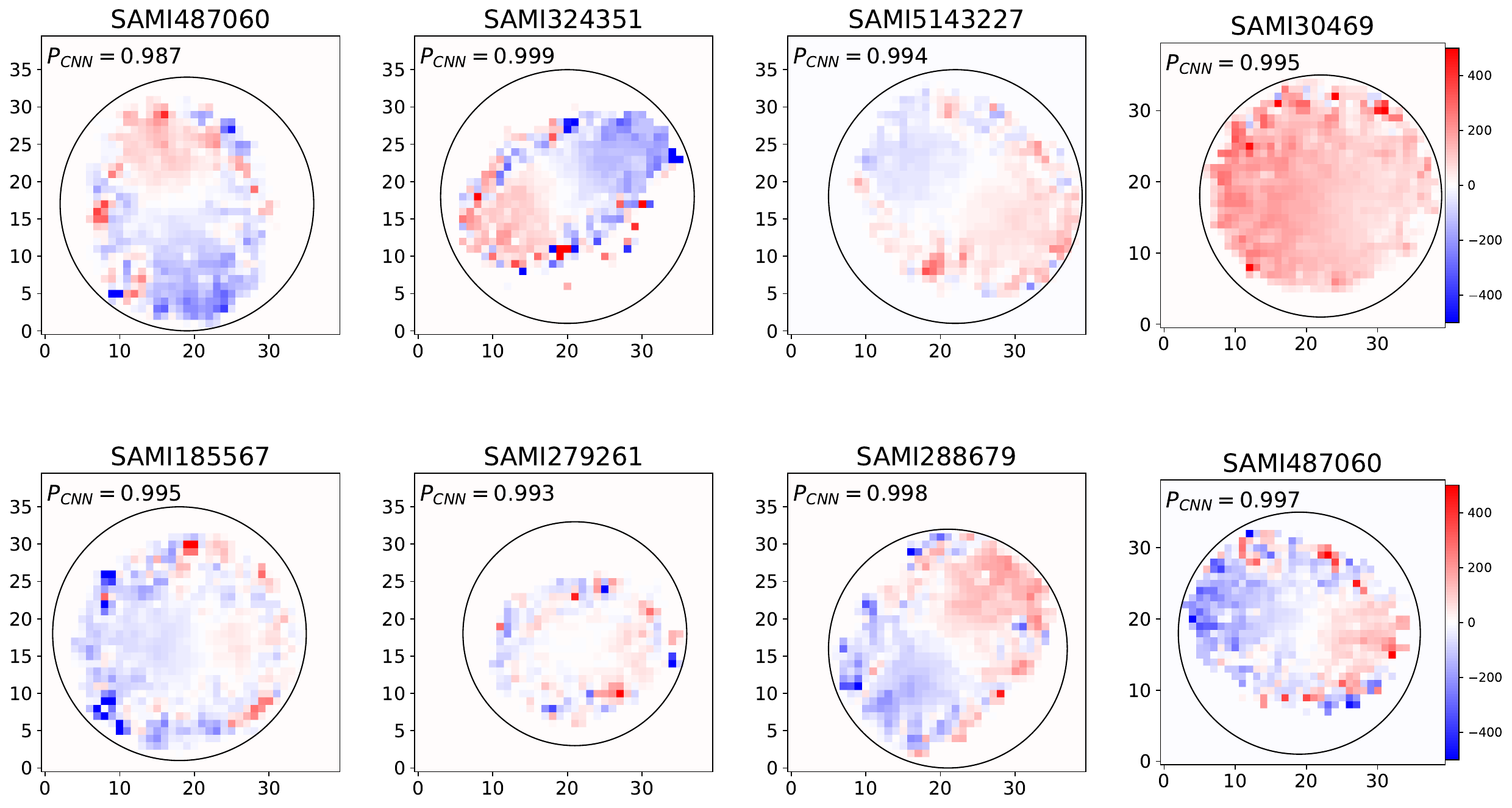}
  
		\caption{These images presents distinct visualizations of false positive samples generated by the CNN model. Each visualization illustrates a different instance where the model mistakenly identified a slow rotator sample as fast one.}
		\label{fp1}
	\end{figure*}

\begin{figure*}[h]
		\centering
		\includegraphics[width=\linewidth]{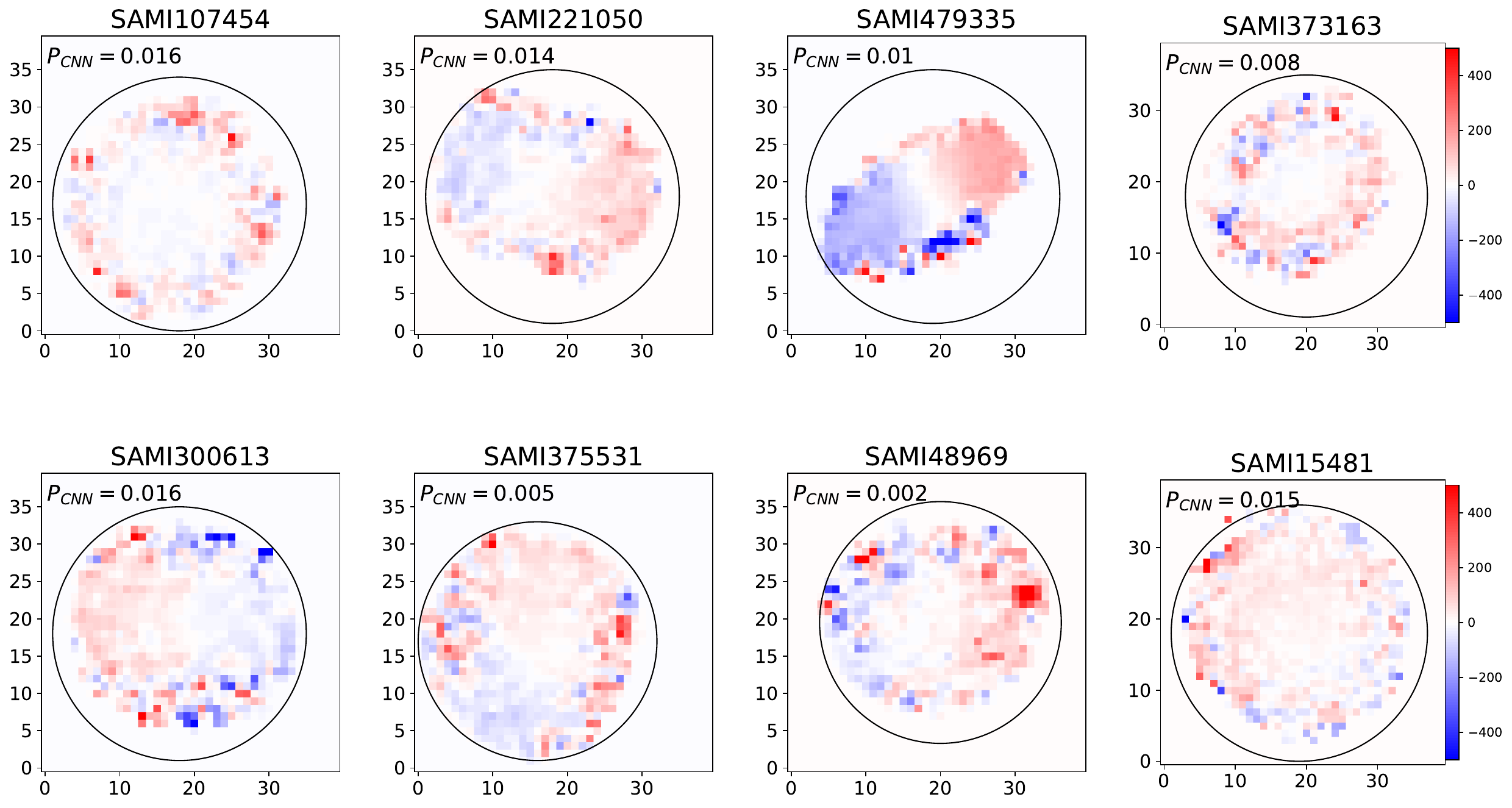}
  
		\caption{These images presents distinct visualizations of false negative samples generated by the CNN model. Each visualization illustrates a different instance where the model mistakenly identified a fast rotator sample as slow one.}
		\label{fn1}
	\end{figure*}

Figure \ref{fp1} displays different visualizations of false positive samples from the CNN model. Each visualization represents a different instance where the model incorrectly classified a slow rotator sample as a fast one. Figure \ref{fn1} displays  visualizations of false negative samples from the CNN model. Each visualization represents a different instance where the model incorrectly classified a fast rotator sample as a slow one.
These false positive and false negative cases are crucial for understanding where and why the model is making mistakes. They could represent edge cases or challenging inputs that share some characteristics with positive samples, leading to misclassification. Analyzing these false positives can provide insights into the model's decision-making process and highlight areas where it might be overfitting or misinterpreting certain features. This information is valuable for refining the model, adjusting its architecture, or enhancing the training data to improve overall performance and reduce such misclassifications in future iterations.
\subsection{Testing the CNN on Unknown Rotators}

\begin{figure*}
		\centering
		\includegraphics[width=\linewidth]{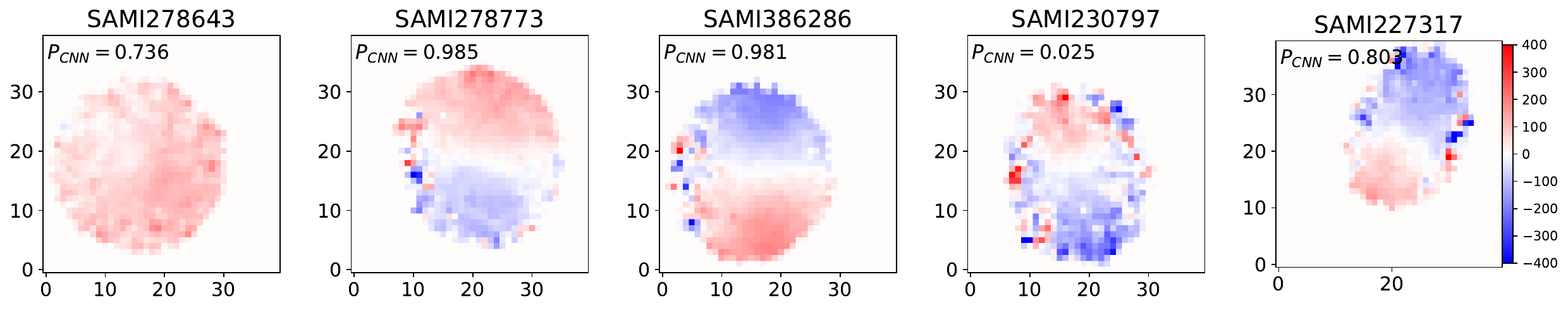}
  \includegraphics[width=\linewidth]{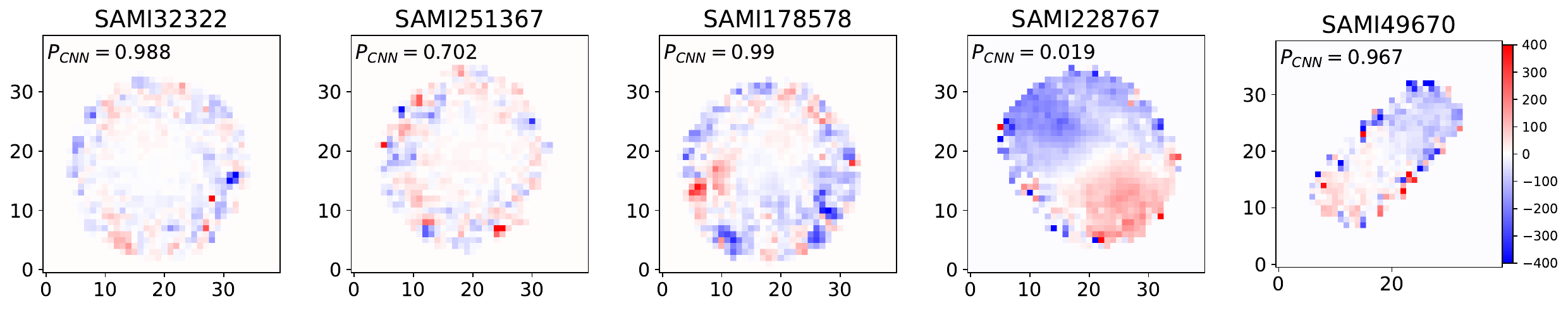}
  \includegraphics[width=\linewidth]{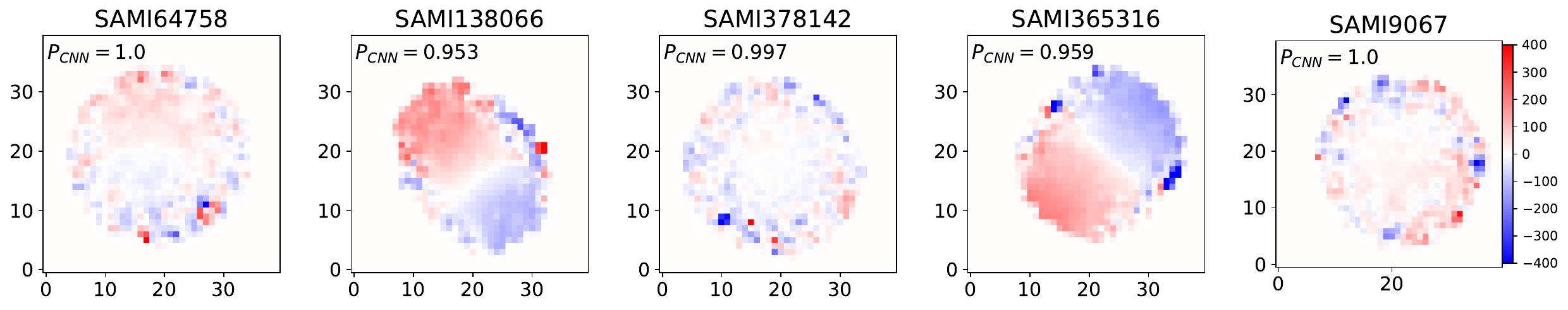}
  \includegraphics[width=\linewidth]{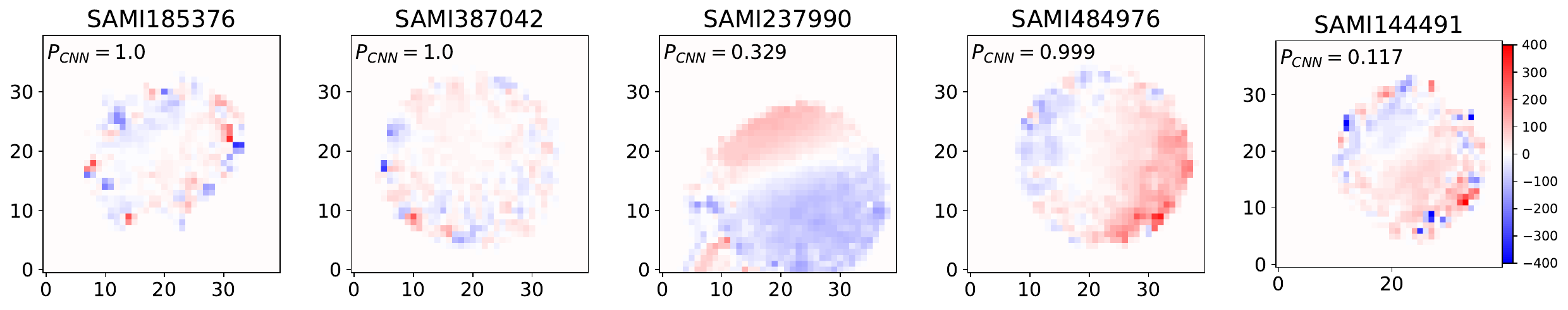}
  \includegraphics[width=\linewidth]{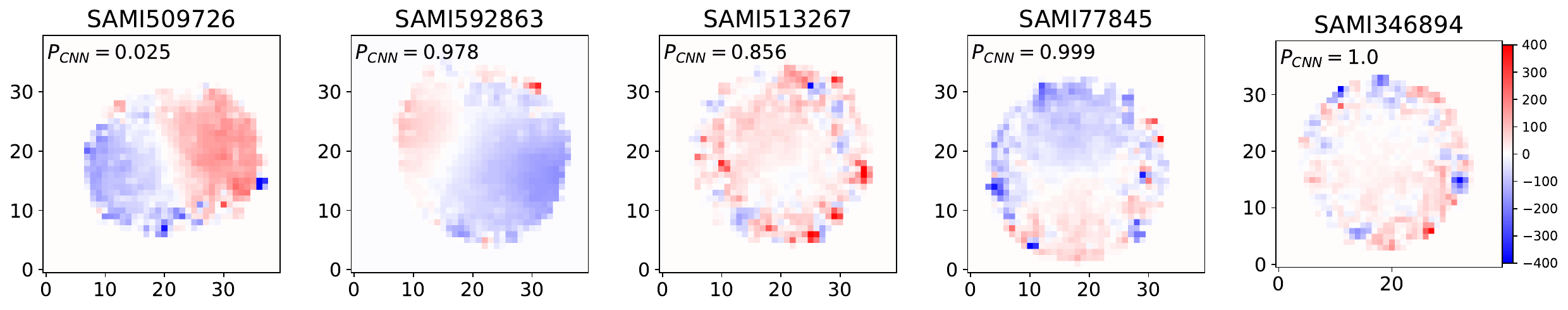}
  \includegraphics[width=\linewidth]{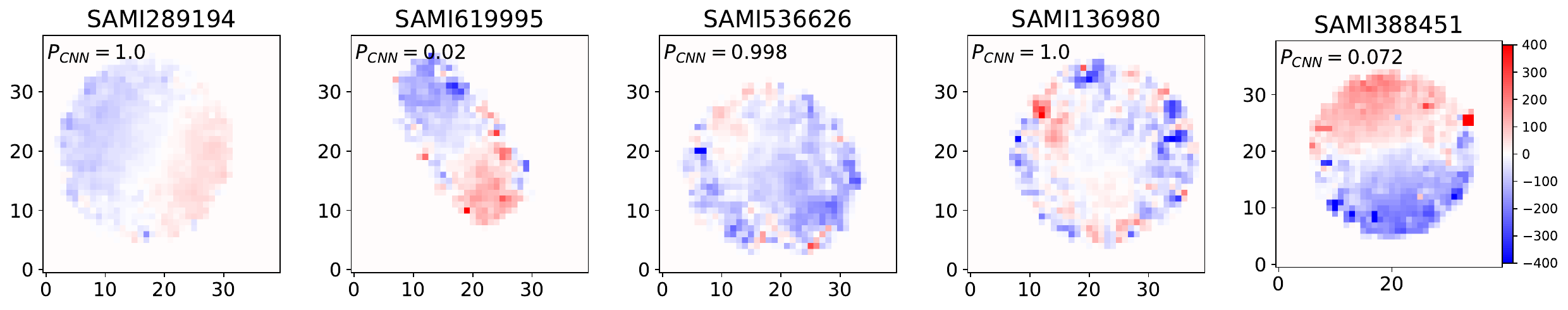}
  \includegraphics[width=\linewidth]{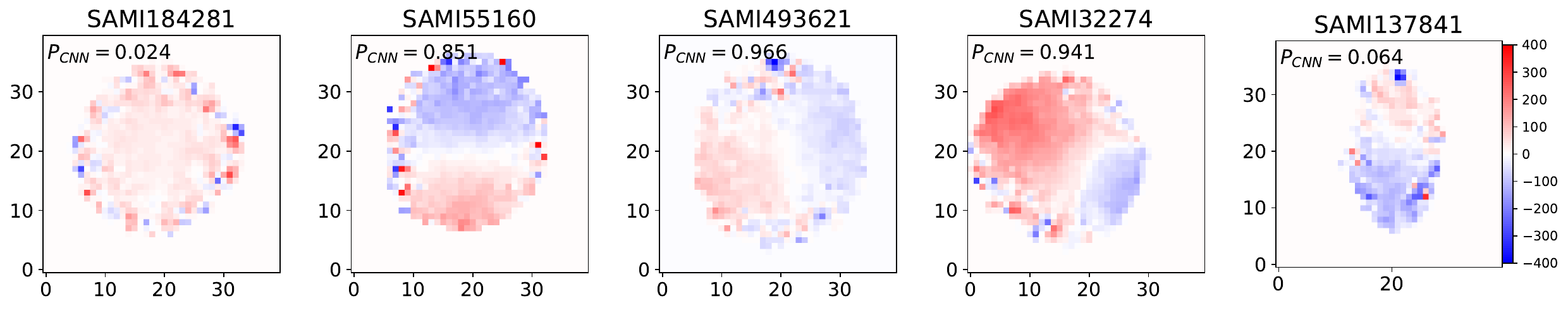}
		\caption{ These images showcases various visualizations of the unknown rotators samples that are face significant challenges in deriving $\lambda_R$-spin parameter. Our CNN model applied on these images and the probability predicted by CNN reported on the top of each image.}
		\label{unknown}
	\end{figure*}

Figure \ref{unknown} illustrates the CNN prediction probabilities for unknown fast and slow rotator samples. Traditional techniques that rely on the $\lambda_R$-spin parameter derived from stellar kinematics often face significant challenges, including the misclassification of galaxies with complex rotation patterns and the inability to effectively analyze low surface brightness galaxies \citep{Harborne2019}. These limitations stem from the conventional methods' struggles to capture the intricate details of stellar motion across diverse galaxy environments. For example, galaxies with non-axisymmetric features or those affected by interactions with neighboring galaxies can yield misleading $\lambda_R$ values. Moreover, low surface brightness galaxies, which often exhibit considerable structural complexity, may fall below the detection thresholds of standard techniques, resulting in gaps in our understanding of their dynamics.

Among the 1,112 samples in the SAMI catalogue, a significant portion, 854 samples were not classified as either slow or fast rotators due to these limitations.
To tackle these complexities, we utilize the strengths of CNNs, which are adept at recognizing patterns in high-dimensional data. By training our CNN on stellar kinematic maps, we can effectively identify subtle variations in rotational dynamics that traditional methods might miss. In Figure \ref{unknown}, we present the probability distribution of our CNN model's predictions for the unknown samples. In this analysis, we consider samples with a probability greater than 0.95 as fast rotators and those with a probability lower than 0.05 as slow rotators. Based on this criterion, our model identifies 678 fast rotators and 39 slow rotators.

The CNN's capability to learn from extensive datasets enables it to discern and classify intricate rotational signatures, thereby enhancing the accuracy of galaxy classification. This methodology not only improves our ability to categorize fast and slow rotators but also allows us to investigate previously ambiguous cases involving complex rotations or low surface brightness galaxies, ultimately advancing our comprehension of galaxy formation and evolution.
   \begin{figure}
		\centering
		\includegraphics[width=1.\linewidth]{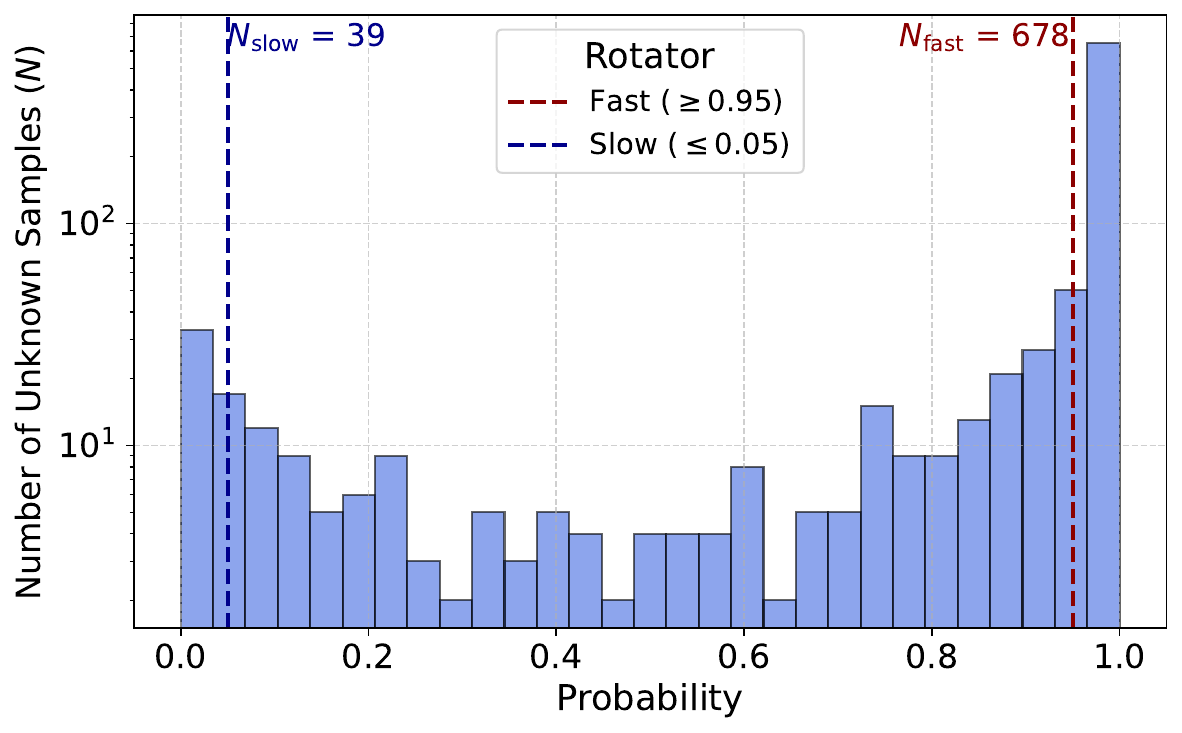}
		\caption{Histogram of CNN prediction probabilities ($P$) for unknown samples. The distribution is marked by two threshold regions: fast rotators ($P \geq 0.95$, red dashed line) and slow samples ($P \leq  0.05$, blue dashed line). The total count of fast ($N_{fast}$) and slow ($N_{slow}$) samples are 678 and 39 respectively.}
		\label{unkown}
	\end{figure}

\subsection{Interpretability of the Model’s Classifications}
This subsection outlines a series of studies aimed at elucidating how our CNN model arrives at its decisions when classifying images. Two primary techniques are used to gain insight into the decision-making processes of the CNN model:

\vspace{0.2cm}
(i) Measuring the clustering of velocity variations in the stellar kinematics maps using Global Moran's I (GMI) tool \citep{moran1948interpretation,chen2013new}.

(ii) Utilizing Integrated Gradients (IG) \citep{sundararajan2017axiomatic} to emphasize the areas of input images that significantly respond to the fast or slow classes.

\vspace{0.2cm}

We analyze the histograms of GMI values for the fast and slow rotators, emphasizing the significance of clustering in high- and low-velocity stars for defining these categories. Afterwards, we employ IG to identify key regions within the velocity stellar maps that influence the decision-making process of our CNN model. Finally, we investigate the model's decisions for true positives and true negatives cases, exploring the relationship between clustering in velocities of stars and the IG values outcomes.

\subsubsection{Clustering of High- and Low- Velocity Stars}
Visual inspection of the velocity variations in the stellar kinematics maps reveals a more pronounced clustering of high- and low-velocity stars in fast rotators compared to slow rotators. In this direction, we utilize the GMI tool to assess the spatial autocorrelation of stellar maps by considering both the locations and values of features simultaneously. Specifically, these observed patterns indicate a clustering of high- and low-velocity stars within the stellar maps.
GMI index value measure the spatial autocorrelation, which quantifies how similar or dissimilar data values are based on their spatial proximity. It’s often used in spatial statistics to determine whether there is clustering, dispersion, or randomness in a data sample. The formula for the GMI index is:

\begin{equation}
I = \frac{n}{\sum_{i=1}^{n} \sum_{j=1}^{n} w_{ij}} \cdot \frac{\sum_{i=1}^{n} \sum_{j=1}^{n} w_{ij} (x_i - \bar{x})(x_j - \bar{x})}{\sum_{i=1}^{n} (x_i - \bar{x})^2},
\end{equation}
where \( n \) is the number of observations, \( x_i \) is the value of the variable of interest at location \( i \), \( \bar{x} \) is the mean of the variable, \( w_{ij} \) is the spatial weight between locations \( i \) and \( j \). The interpretation of the GMI index value can be categorized as follows:

\vspace{0.2cm}

 (i): Positive GMI index (close to +1): Indicates clustering or positive spatial autocorrelation, meaning similar values (either high or low) tend to be close to each other.
 
(ii): Negative GMI index (close to -1): Indicates dispersion or negative spatial autocorrelation, suggesting that high and low values are intermixed and tend to avoid clustering.

(iii): Zero or near-zero GMI index: Suggests a random spatial pattern with no clear clustering or dispersion.

\begin{figure}
		\centering
		\includegraphics[width=1.\linewidth]{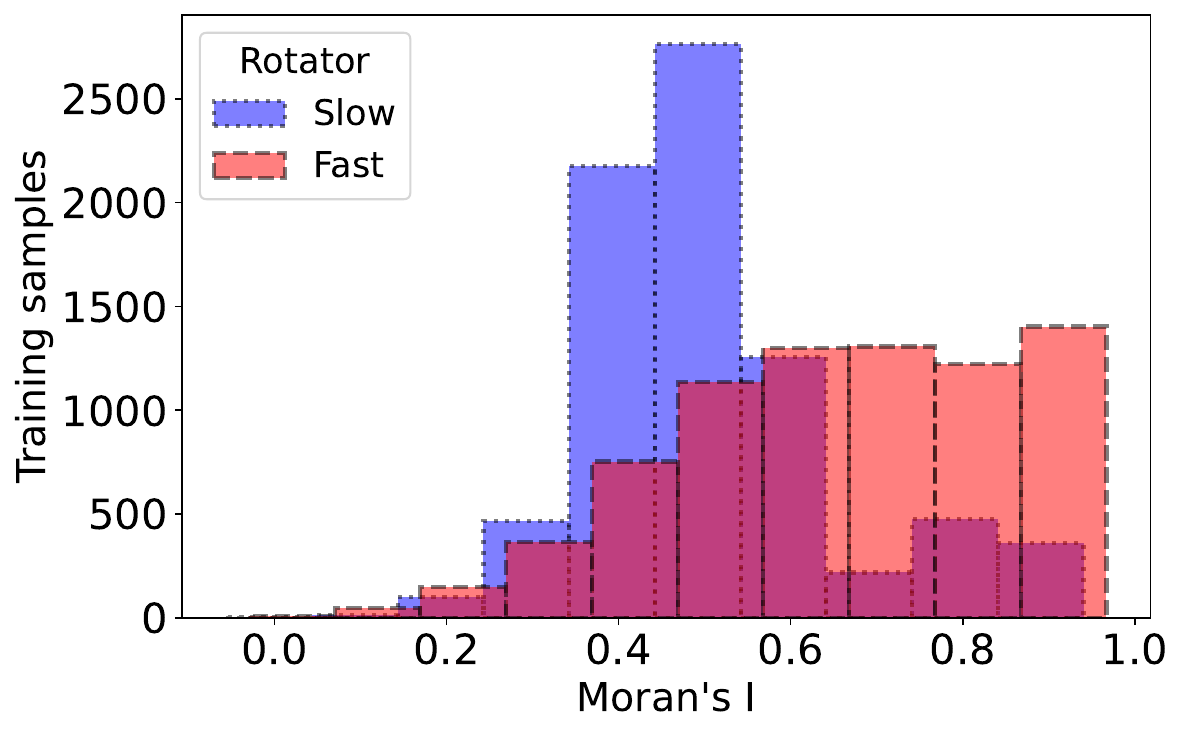}
		\caption{Distribution of Moran's I for slow and fast rotators, illustrating differences in spatial autocorrelation. Slow rotators (blue) exhibit a narrower distribution centered around lower values, while fast rotators (red) display a broader distribution at higher values.}
		\label{prob1}
	\end{figure}

\begin{figure*}
		\centering
		\includegraphics[width=\linewidth]{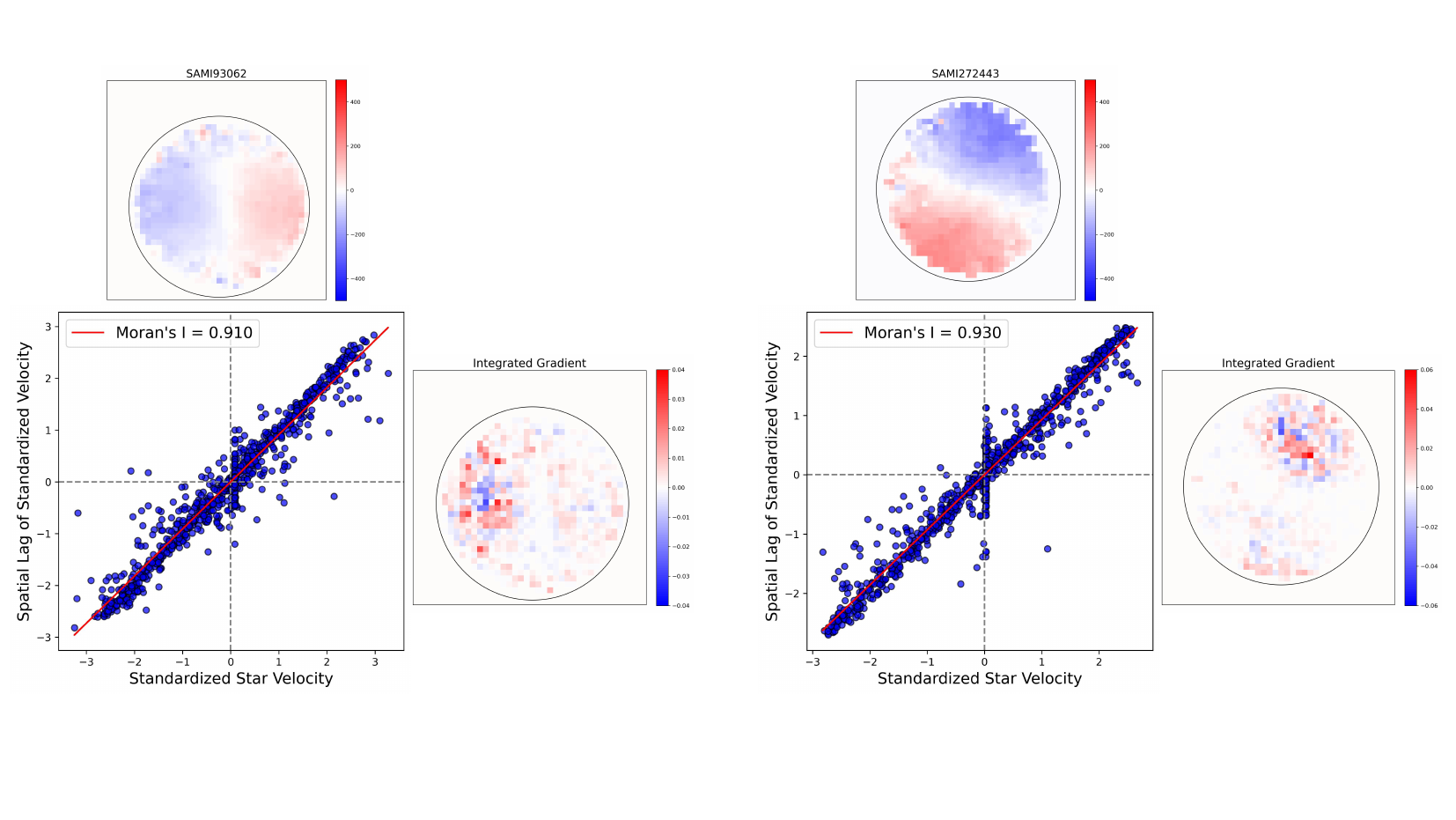}
  
		\caption{Spatial autocorrelation of star velocities for two fast rotators from the SAMI dataset predicted by our CNN model. The central scatter plot shows the Moran scatter plot for each fast rotator, where the spatial lag of standardized velocity is plotted against the standardized star velocity. The top row presents spatial velocity maps, with red and blue regions representing higher and lower velocities, respectively. The right row displays the Integrated Gradients (IG) values for each pixel, where positive IG values contribute positively to the model's decision. }
		\label{fast_moran}
	\end{figure*}

\begin{figure*}
		\centering
		\includegraphics[width=\linewidth]{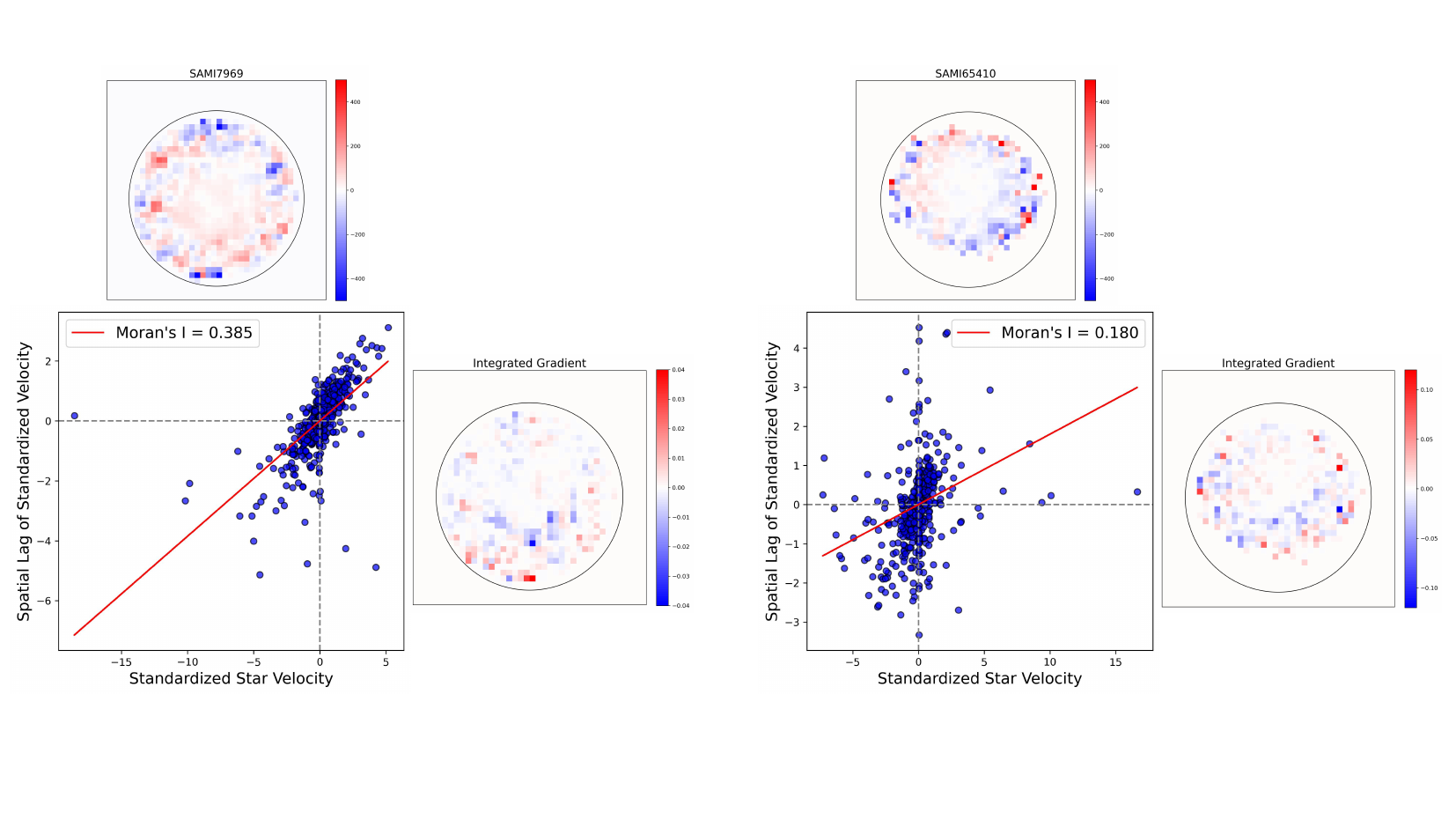}
  
		\caption{Spatial autocorrelation of star velocities for two slow rotators from the SAMI dataset. The central scatter plot shows the Moran scatter plot for each slow rotator, where the spatial lag of standardized velocity is plotted against the standardized star velocity. Lower Moran's I values (0.385 and 0.180) suggest weaker spatial autocorrelation, indicating less coherent kinematic structures compared to the fast rotators analyzed in Figure \ref{fast_moran}. The bottom row displays the Integrated Gradients (IG) values for each pixel, highlighting that the majority of IG values are concentrated in regions with weaker spatial autocorrelation, further indicating less coherent kinematic structures in the slow rotators.}
		\label{slow_moran}
	\end{figure*}

Figure \ref{prob1} shows the histograms of GMI values for
fast and slow rotators. The distribution shows that there are significantly more slow rotators in the lower ranges (0.0 to 0.4) of Moran's I. The fast rotators more prominently in the middle to higher ranges (approximately 0.4 to 1.0), suggesting a different distribution pattern between the two categories. In figure \ref{fast_moran} we show the 
spatial autocorrelation in star velocities maps for two fast rotators from the SAMI dataset which is predicted correctly by our CNN model. The central scatter plot presents Moran scatter plot for each fast rotator, where the spatial lag of standardized velocity is plotted against the standardized star velocity. High Moran's I values (0.910, and 0.930) indicate strong spatial autocorrelation in all cases, suggesting coherent kinematic structures within these fast rotators.  The top row shows spatial velocity maps, with red and blue regions indicating higher and lower velocities, respectively. In contrast, in figure \ref{slow_moran} we show the previous analysis for the two slow rotators.  Lower Moran's I values (0.385, and 0.180) suggest weaker spatial autocorrelation, indicating less coherent kinematic structures compared to the fast rotating galaxies.

The examination of stellar kinematics and the Global Moran's I (GMI) index reveals substantial differences between fast and slow rotators in terms of velocity distribution. Fast rotators exhibit more distinct clustering of both high and low velocities, with GMI values between 0.4 and 1.0, indicating a higher degree of spatial autocorrelation and more uniform kinematic structures. Conversely, slow rotators display lower GMI values (0.0 to 0.4),  pointing to weaker spatial autocorrelation and less coherent velocity distributions. This suggests that fast rotators have more cohesive velocity arrangements than slow rotators, which exhibit more scattered and irregular distributions.

\subsubsection{Integrated Gradients}
Integrated Gradient (IG) \citep{sundararajan2017axiomatic}  is a systematic method used to evaluate the contribution of individual pixels or regions in an input image to the model's final classification decision. To thoroughly understand IG, it is crucial to explore the concept of alpha, a key parameter in this approach. Alpha, is a real number ranging from 0 to 1. It establishes a path along which the gradients of the model's predicted probabilities are evaluated. Specifically, $\alpha$ dictates the proportional contribution of information from a baseline image relative to the target image being analysed. This proportional contribution is calculated as an integral along the defined path. Recently IGs has numerous applications in astronomy, including strong lensing searches \citep{wilde2022detecting} and weak lensing analysis \citep{matilla2020interpreting}.

IG is applied on the slow and fast rotators in the figures \ref{fast_moran}, \ref{slow_moran}, respectively. In Figure \ref{fast_moran}, in the right row of each sample, we show the IG value for each pixels of these samples. It should be noted that pixels with IG values higher than zero have positive effect on the our model decision and vice versa. The majority of these IG images highlight the pixels in the coherent kinematic structures in the image. So according to this analysis we can say one of the the CNN model features that is used in distinguishing fast and slow rotators is the coherent kinematic structures in the fast rotators. While, in Figure \ref{slow_moran}, slow rotators IG images the majority of pixels are concentration in the weaker spatial autocorrelation, indicating less coherent kinematic structures.

\section{Discussion and Conclusion}\label{sec4}
In this work, we have demonstrated the success of convolutional neural networks (CNNs) in effectively classifying galaxies into fast and slow rotators based on their stellar kinematic maps. Using data from the SAMI survey, we developed and fine-tuned a CNN architecture that delivered an accuracy of 91\% and a precision of 95\%, proving its reliability in distinguishing between the two categories.  These results illustrate the the potential of the CNNs to overcome challenges in galaxy classification,
particularly when traditional methods relying on $\lambda_R$-spin parameters, encounter uncertainties.

The analysis of stellar kinematics using the Global Moran's I (GMI) index reveals distinct differences between fast and slow rotators in their velocity distributions. This suggests that fast rotators have more organized and structured velocity fields, whereas slow rotators exhibit more scattered and irregular distributions, highlighting fundamental differences in the kinematic properties of slow and fast rotators. The interpretability of CNN classifications is a important component of this study. By applying Integrated Gradients, we uncovered the primary kinematic feature that  that significantly effected the model’s outputs. Our findings indicate that the CNN focuses on coherent kinematic structures within stellar velocity maps to classify galaxies, confirming its ability to detect meaningful physical patterns. Using this interpretable CNN model we can enhance confidence in the model’s predictions.

The scalability and adaptability of CNNs are essential given the emergence of current and future large-scale integral field spectroscopy (IFS) surveys, such as the Hector instrument \citep{bryant2020hector, bryant2024hector} which is designed to characterize 15,000 galaxies and Wide-field Spectroscopic Telescope (WST) \cite{Mainieri2024}, which generate datasets comprising billions of galaxies. Under this condition, traditional statistical approaches will face significant challenges in handling such extensive data effectively.  Our findings highlight the vital role of CNNs for processing and analysing stellar kinematic patterns in these large datasets, establishing their importance for the next generation of astronomical exploration. 

Although the findings are promising, the study highlights some limitations. The model showed uncertainty in borderline cases, as shown in the probability distribution of predictions.  Addressing this will need for larger and more diverse training datasets. Future research could aim to include expanding the approach to account for advanced kinematic features, like higher-order velocity moments, or leveraging transfer learning with pre-trained models on similar datasets. Furthermore, employing more explainable CNN models allows us to identify the most significant features in the decision-making process. This understanding can be leveraged to enhance the model's accuracy by incorporating attention layers \citep{cordonnier2019relationship}, which focus on the most relevant aspects of the data. We intend to broaden our analysis by investigating CNN alongside traditional machine learning classifiers, including random forests \citep{breiman2001random}, support vector machines \citep{suthaharan2016support}, and Naive Bayes \citep{webb2010naive}, with a focus on both accuracy and model interpretability.


\vspace{6pt} 





\authorcontributions{The first two authors are considered the main contributors to this paper and are listed as co-first authors, acknowledged by first name. Conceptualization: A.CH. and F.F.; Methodology: A.CH. and F.F.; Software: M.R.; Validation: M.R., B.F., and F.V.; Formal Analysis: A.CH. and F.F.; Investigation: F.F.; Resources: M.R.; Data Curation: M.R.; Writing—Original Draft Preparation: M.R.; Writing—Review and Editing: M.R.; Visualization: A.CH.; Supervision: M.R. and F.V. and B.F; Project Administration: A.CH. and F.F.; Funding Acquisition: F.F. All authors have read and agreed to the published version of the manuscript. , please turn to the  \href{http://img.mdpi.org/data/contributor-role-instruction.pdf}{CRediT taxonomy} for the term explanation. Authorship must be limited to those who have contributed substantially to the work~reported.}

\funding{This research received no external funding}

\dataavailability{The codes used in this study are publicly available on GitHub at the following repository: \url{https://github.com/AM-Chegeni/galaxy_kinematics_CNN}. The data underlying this article will be shared on reasonable request to the corresponding author. The SAMI data presented in this paper are available from Astronomical Optics’ Data Central service at: https://datacentral.org.au/.}

\acknowledgments{F.F and M.R would like to express her heartfelt appreciation to EuroSpaceHub and LUNEX EuroMoonMars Earth Space Innovation for their generous funding and unwavering support. A.CH. was supported by the MUR
PRIN2022 project 20222JBEKN with title "LaScaLa" - funded by the European Union - NextGenerationEU. }

\conflictsofinterest{The authors declare no conflicts of interest.}






\isPreprints{}{ This command is only used for ``preprints''.
\begin{adjustwidth}{-\extralength}{0cm}
} 
\printendnotes[custom] 

\reftitle{References}


\bibliography{output.bib}


%


\PublishersNote{}
\isPreprints{}{
\end{adjustwidth}
} 
\end{document}